\documentclass[aps,pra,10pt,twocolumn,floatfix,showpacs,numerical,footinbib,longbibliography]{revtex4-1}
\usepackage{amsmath}
\usepackage{amssymb}
\usepackage{dsfont}
\usepackage{bbold}
\usepackage[%
colorlinks=true,
urlcolor=blue,
linkcolor=blue,
citecolor=blue]{hyperref}
\usepackage{tikz}
\usepackage{graphicx}
\usepackage{subfigure}
\usepackage{svg}
\usepackage{mathtools}
\usepackage[utf8]{inputenc}

\begin{document}
	\renewcommand{\thefootnote}{\fnsymbol {footnote}}
	
	\title{\textbf{Dynamical redistribution of quantum resources in tree-level Bhabha scattering}}
	\author{Zan Cao}
	\author{Meng-Long Song}
	\author{Xue-Ke Song}
	\author{Liu Ye}
	\author{Dong Wang}
	\email{dwang@ahu.edu.cn}
	\affiliation{State Key Laboratory of Opto-Electronic Information Acquisition and Protection Technology, School of Physics, \textcolor{blue}{Anhui University}, Hefei 230601,  China}
	\begin{abstract} 
The fundamental interactions governed by quantum electrodynamics (QED) are intrinsically rich in quantum resources, yet how these resources dynamically redistribute during relativistic scattering is still not fully understood. In this work, we systematically investigate the tree-level Bhabha scattering process ($e^- e^+ \rightarrow e^- e^+$) within the framework of quantum resource theory, revealing how QED kinematics and Feynman amplitudes strictly dictate resource redistribution. Specifically, we demonstrate a strict anti-correlation between entropic uncertainty and dynamically generated entanglement across diverse initial states. We find that mass-induced single-helicity-flip transitions cause a pronounced geometric symmetry breaking in the non-relativistic regime, whereas the restoration of chiral symmetry in the ultra-relativistic limit ensures strict symmetry about the backward scattering angle. Furthermore, we analytically establish a rigorous equivalence between local wave-particle duality and global bipartite quantum coherence. Finally, evaluating the trade-off between local duality and Bell nonlocality, we show that in the ultra-relativistic limit, transverse scattering  of basic factorized states equalizes the $s$- and $t$-channel amplitudes to optimize non-local correlations. However, pre-existing local coherence inevitably disrupts this delicate kinematic balance, significantly suppressing the Bell parameter and preventing the maximal violation of local realism. Therefore, we believe the present results provide deeper understanding of the fundamental quantum nature of QED processes.
	\end{abstract}
	\maketitle

	\section{Introduction}
   In recent years, the intersection of high-energy physics and quantum information science has emerged as a vibrant frontier for exploring the foundational principles of quantum mechanics. High-energy environments, characterized by relativistic scattering states, provide unique platforms to investigate non-classical phenomena that are otherwise inaccessible in low-energy systems \cite{1.1, 1.2, 1.3,11.11JHEP,11.12PhysRevD.108.025015,11.13PhysRevC.108.L041601,11.14PhysRevD.104.116021,11.15PESCHANSKI201689,11.16PhysRevD.107.116007,11.17PhysRevLett.125.181602,11.18PhysRevD.95.114008,2.7spectatorPhysRevD.109.096022}. Within this context, substantial theoretical and experimental efforts have been dedicated to certifying quantum signatures across various high-energy regimes. For instance, quantum entanglement and diverse quantum correlations have been extensively studied in neutrino oscillations \cite{1.4,1.5,1.6,1.7,1.8,1.9,1.10,1.11,1.12,1.13,1.14,1.15} and top-quark pair production at the Large Hadron Collider (LHC) \cite{1.16Afik2022quantuminformation,1.17PhysRevLett.130.221801,1.18fhkc-kfhr,1.19CHEN2026140426}. Furthermore, other fundamental aspects of quantum mechanics, such as quantum coherence and entropic uncertainty relations, are attracting growing interest within these same relativistic environments \cite{1.19Relativistic2024,1.20PhysRevD.110.055025,1.21WANG2024129364,1.22WANG2024138876}.
   Collectively, these investigations demonstrate that relativistic scattering and decay processes are inherently rich in quantum resources, establishing a solid foundation for probing quantum field theory behaviors through an information-theoretic lens.
   

   To formalize the characterization of these multifaceted non-classical phenomena, quantum resource theory (QRT) provides a rigorous and unified mathematical framework \cite{2.1RevModPhys.91.025001}. Rather than treating fundamental features—such as entanglement, quantum coherence, and entropic uncertainty \cite{2.1.1RevModPhys.81.865,2.1.2RevModPhys.89.041003,2.1.3RevModPhys.89.015002}—as isolated kinematic traits, this modern framework conceptualizes them as quantifiable physical resources. Within this paradigm, these resources are governed by strict operational constraints, conservation rules, and dynamic conversion laws \cite{2.1.4CHEN2026140426}. Consequently, one of the central objectives of QRT is to formulate quantitative trade-off and complete complementarity relations. These relations are crucial, as they delineate the permissible parameter spaces of quantum states through complementary observables, thereby setting fundamental boundaries on the simultaneous existence of incompatible quantum properties \cite{2.2Fanxiaogang2019,2.3Mingfei2021,2.4PhysRevA.106.042415}. For instance, established theoretical frameworks have successfully mapped the exact resource conversion dynamics between local coherence and bipartite correlations \cite{2.6PhysRevLett.123.220501}. These studies reveal a strict dynamical trade-off, demonstrating how the consumption of local superposition resources can be operationally quantified to fuel the generation of non-local entanglement.
   
   Despite these conceptual advances, most investigations into quantum resource interconversion have predominantly focused on non-relativistic systems or phenomenological bipartite models. Recently, a pioneering study by Blasone et al. \cite{3.1TreelevelQRDPhysRevD.111.016007} took a significant step forward in this field by systematically evaluating the complete complementarity relations (CCR) in tree-level quantum electrodynamics (QED) processes, revealing the generation and distribution of entanglement among helicity degrees of freedom. However, while CCR provides a powerful triality relation among predictability, visibility,  and concurrence, the dynamic interplay between QED kinematics and a broader spectrum of quantum features—such as entropic uncertainty, bipartite duality, coherence, and Bell nonlocality—remains largely unexplored. In this work, we adopt the Bhabha scattering framework ($e^-e^+ \rightarrow e^-e^+$) established in \cite{3.1TreelevelQRDPhysRevD.111.016007} and significantly extend this resource-theoretic paradigm. By naturally incorporating both $s$-channel annihilation and $t$-channel scattering pathways, this fundamental QED process provides a clean mathematical framework.  This allows us to precisely evaluate how intrinsic physical properties—specifically finite fermion masses and the recovery of chiral symmetry—strictly dictate the complex interplay and interconversion among these diverse quantum resources.
   
   The remainder of this paper is organized as follows. In Sec. \ref{II}, we briefly review the theoretical preliminaries of the Bhabha scattering process and the relevant quantum resources, including entropic uncertainty relations, the complete complementarity relations balancing wave-particle duality and entanglement, quantum coherence, and Bell nonlocality. In Sec. \ref{III}, we investigate the dynamical generation of quantum entanglement and its strict geometric anti-correlation with the entropic uncertainty bound. In Sec. \ref{IV}, we analytically establish the fundamental equivalence between local wave-particle duality and global bipartite quantum coherence. Building upon this, then, the connection between duality and the maximal violation of Bell inequalities is examined in Sec. \ref{V}. Finally, a brief conclusion is presented in Sec. \ref{VI}.
 
	\section{SETTING AND PRELIMINARIES}\label{II}
	\subsection{Bhabha scattering process}\label{IIA}
	In this section, the formalism established in Ref. \cite{3.1TreelevelQRDPhysRevD.111.016007} is briefly reviewed, where a general tree-level Bhabha process involving electron-positron scattering ($e^-e^+\rightarrow e^-e^+$) is examined. 
	For simplicity, we carry out the calculations in the center-of-mass (COM) frame of the electron-positron pair.
	The internal product of fermion states is defined as
	\begin{equation}
		\langle k, a \mid p, b\rangle=2 E_{\mathbf{k}}(2 \pi)^{3} \delta^{(3)}(\mathbf{k}-\mathbf{p}) \delta_{a, b},
		\label{Eq.1}
	\end{equation}
	where $k$ and $p$ are the 4-momenta, $a$ and $b$ are the spin indices.
	
	For a given initial state, the final state after Bhabha scattering is a superposition of possible outcomes, each outcome's probability given by the scattering amplitude.
	$\mathcal{M}\left ( p_{1}, a, p_{2}, b; p_{3}, r, p_{4}, s \right ) $. As an illustrative case, we consider a generic separable initial state
	\begin{equation}
		\left|i\right\rangle = \left|p_1, a\right\rangle_{A} \otimes \left|p_2, b\right\rangle_{B},
		\label{Eq.2}
	\end{equation} 
	the scattering matrix (S-matrix) is defined as $S=\mathbb{I}+iT$, where $\mathbb{I}$ is the identity operator for non-scattering processes, and $T$ is the transition operator describing QED-mediated Bhabha scattering — the core process of this study. After scattering, the final state is
	\begin{equation}
		\begin{split}  	  
			|f\rangle = |i\rangle &+ i\sum_{r,s}\int \frac{d^3\mathbf{p}_3 d^3\mathbf{p}_4}{(2\pi)^6 2E_{\mathbf{p}_3} 2E_{\mathbf{p}_4}} \\
			&\times \delta^{(4)}(p_1 + p_2 - p_3 - p_4) \\
			&\times \left[\mathcal{M}(a,b;r,s)|p_3,r\rangle_A \otimes |p_4,s\rangle_B\right],
			\label{Eq.3}
		\end{split}
	\end{equation}
	where, $\mathcal{M}(a,b;r,s)$ denotes the scattering amplitude, here only the spin indices of the initial and final states are retained for simplicity. The system's density operator can  be derived from the above expression
	\begin{equation}
		\rho_{}^f=\frac{1}{\mathcal{N} } |f \rangle \langle f|,
		\label{Eq.4}
	\end{equation}
	where, $\mathcal{N}$ is the normalization constant. Taking the partial trace over the subsystems yields
	\begin{equation}
		\text{Tr}_X[\rho_{}^f]=\sum_\sigma\int\frac{d^3\mathbf{k}}{(2\pi)^32E_\mathbf{k}}(\mathbb{I}  _r\otimes_X\langle k,\sigma|)\rho_{}^f(\mathbb{I}  _r\otimes|k,\sigma\rangle_X),
		\label{Eq.5}
	\end{equation}
	where $ \mathbb{I}_r$ is the identity operator on the residual subspaces, $k$ and $ \sigma $ denote the 4-momentum and spin indices respectively, and $X$ defines the generic space for trace computation. Employing Eq. (\ref{Eq.1}) and the relations
	\begin{equation}
		2\pi \delta^{(0)}(E_i - E_f) = \int_{-T/2}^{T/2} e^{i(E_i - E_f)t} dt,
		\label{Eq.6}
	\end{equation}
	\begin{equation}
		(2\pi)^3 \delta^{(3)}(\mathbf{k} - \mathbf{p}) = V \delta_{\mathbf{k},\mathbf{p}},
		\label{Eq.7}
	\end{equation}
	when $E_i = E_f$, $\mathbf{k} =\mathbf{p}$, we can obtain $(2\pi) \delta^{(0)}(0) = T $ and $ (2\pi)^3 \delta^{(3)}(0) = V$. The normalization constant $\mathcal{N}$ can be readily computed as
	\begin{equation}
		\mathcal{N} = \text{Tr}_A[\text{Tr}_B[|f\rangle\langle f|]].
		\label{Eq.8}
	\end{equation}
	
	\subsection{\textbf{Entropic uncertainty relations}}\label{IIB}
	The uncertainty principle is a fundamental cornerstone of quantum mechanics. In quantum information theory, it is rigorously formulated using Shannon or von Neumann entropies. For bipartite systems, the presence of quantum entanglement fundamentally alters the uncertainty bounds. Renes and Boileau \cite{En1.1Renes.Boileau2009}, followed by Berta et al. \cite{En1.2Berta2010}, pioneered the quantum-memory-assisted entropic uncertainty relation (QMA-EUR). If two incompatible observables $\hat{Q}$ and $\hat{R}$ are measured on subsystem $A$, while subsystem $B$ serves as a quantum memory, the uncertainty is bounded by
		\begin{equation}
			S(\hat{Q}|B) + S(\hat{R}|B) \ge S(A|B) - \log_2 c(\hat{Q},\hat{R}),
			\label{Eq.9}
		\end{equation}
		where $c(\hat{Q},\hat{R}) \equiv \max_{ij} |\langle \psi_i | \phi_j \rangle|^2$ represents the maximum overlap between the eigenvectors $|\psi_i\rangle$ of $\hat{Q}$ and $|\phi_j\rangle$ of $\hat{R}$. The term $S(\hat{Q}|B) = S(\rho_{\hat{Q}B}) - S(\rho_{B})$ defines the conditional von Neumann entropy of the system after $\hat{Q}$ measurement on particle $A$, and the state of the compound system after measurement can be expressed as 
		$\rho_{\hat{Q}B} = \sum_{i} \left( | \psi_i \rangle \langle \psi_i | \otimes \mathbb{I}_{B} \right) \rho_{AB} \left( | \psi_i \rangle \langle \psi_i | \otimes \mathbb{I}_{B} \right)$, where $|\psi_i\rangle$ is the eigenvector of observable $\hat{Q}$ and $\mathbb{I}_{B}$ is the identity matrix in the Hilbert space of $B$. 
		Similarly for $S(\hat{R} | B)$ and $\rho_{\hat{R}B}$.
	 $S(A|B) = S(\rho_{AB}) - S(\rho_B)$ is the conditional entropy of the pre-measurement bipartite state. Operationally, this relation can be understood through an information-theoretic game. In the context of our QED scattering framework, the outgoing electron acts as the primary system (Alice), while the dynamically entangled outgoing positron serves as the quantum memory (Bob). The QMA-EUR dictates that the more entangled the $e^- e^+$ pair becomes during the scattering process (resulting in a negative $S(A|B)$), the more accurately the positron can predict the measurement outcomes of the electron's non-commuting observables, effectively lowering the uncertainty bound.
	\subsection{\textbf{Complete complementary relationship}}\label{IIC}
	In fundamental quantum systems, wave-particle duality is intrinsically constrained by the entanglement shared with other subsystems. The quantitative trade-off between the particle-like predictability $P$ and the wave-like visibility $V$ was initially formulated by Wootters and Zurek as the inequality $P^2 + V^2 \le 1$ \cite{CCR1.Wootters1979}. For pure bipartite states, this inequality saturates and can be generalized into a complete complementarity relation (CCR), often referred to as the triality relation \cite{CCR2.PhysRevLett.98.250501}
	\begin{equation}
			P_k^2 + V_k^2 + C^2 = 1,
			\label{Eq.10}
	\end{equation}
	where $k \in \{A, B\}$ denotes the respective subsystem.  Here, $C$ represents the systemic entanglement, quantified by the concurrence \cite{CCR3.PhysRevLett.80.2245}
	\begin{equation}
		C(\psi) = |\langle\psi| (\sigma_y \otimes \sigma_y)|\psi^*\rangle|.
		\label{Eq.11}
	\end{equation}
	The predictability $P_k$ and visibility $V_k$ quantify the local which-path information and interference capability, respectively. Operationally, they are defined via the local Pauli operators
	\begin{equation}
		\begin{split}
			P_k = |\langle\psi|\sigma_{z,k}|\psi\rangle|,\quad V_k = 2|\langle\psi|\sigma_{k}^+|\psi\rangle|.
			\label{Eq.12}
		\end{split}
	\end{equation}
	 In this framework, this relation ensures that the dynamical generation of entanglement $C$ between the outgoing electron and positron during Bhabha scattering strictly bounds their local chiral duality behaviors.
	\subsection{Quantum coherence}\label{IID}
	Quantum coherence is a fundamental resource that formally characterizes the capacity of a quantum system to exhibit interference phenomena \cite{XG.1PhysRevA.92.022112}. In bipartite scenarios, such as the electron-positron pairs emerging from QED scattering, the degree of systemic coherence can be robustly quantified through the local states of the subsystems. For two-qubit systems, a well-established measure for the bipartite coherence is formulated as \cite{XG.2RevealingHidden20155}
\begin{equation}
	D_{AB}=\sqrt{\frac{D_{A}^{2}+D_{B}^{2}}{2}},
	\label{Eq.13}
\end{equation}
	where $D_{A} = \sqrt{2\text{Tr}(\rho_{A}^{2})-1}$ and $D_{B} = \sqrt{2\text{Tr}(\rho_{B}^{2})-1}$ denote the local coherence of subsystems $A$ and $B$, respectively. Here, $\rho_{A(B)}$ represents the reduced density matrix obtained by tracing over the complementary subsystem. Crucially, by defining coherence through the purity of the reduced local states, this formulation provides a direct mathematical conduit to link global bipartite quantumness with local wave-particle duality parameters.
\subsection{Bell nonlocality}\label{IIE}
	Bell nonlocality represents a stringent form of quantum correlation that strictly precludes any description by local hidden variable (LHV) models \cite{Bell1.PhysRevLett.98.140402,Bell2.PhysicsPhysiqueFizika.1.195}. For bipartite systems, such as the emergent electron-positron pairs in our scattering framework, this nonlocality is standardly detected via the maximal violation of the Bell-CHSH inequality. Given two dichotomic observables $\vec{a}\cdot\vec{\sigma}$ and $\vec{a}'\cdot\vec{\sigma}$ for particle A, and $\vec{b}\cdot\vec{\sigma}$ and $\vec{b}'\cdot\vec{\sigma}$ for particle B, the CHSH operator is defined as \cite{Bell3.Clauser.PhysRevLett.23.880}
\begin{equation}
	\hat{B} = \vec{a}\cdot\vec{\sigma} \otimes (\vec{b}+\vec{b}')\cdot\vec{\sigma} + \vec{a}'\cdot\vec{\sigma} \otimes (\vec{b}-\vec{b}')\cdot\vec{\sigma}.
	\label{Eq.14}
\end{equation}
	According to LHV models, the expectation value is strictly bounded by $|\langle\hat{B}\rangle|\le2$. A violation of this bound ($|\langle\hat{B}\rangle|>2$) unequivocally certifies the presence of Bell-nonlocal correlations. 
	Crucially, for arbitrary bipartite pure states in the helicity space generated via unitary QED evolution, the maximal violation of the Bell-CHSH inequality is analytically determined by the systemic entanglement.
	This exact mapping provides a powerful tool to translate dynamic entanglement features directly into observable non-local signatures.

\section{ENTROPY UNCERTAINTY AND ENTANGLEMENT IN BHABHA SCATTERING}\label{III}
\subsection{Basis factorized states}\label{IIIA}
We initialize the bipartite system with a two-particle separable state, for example, a right-handed electron and a left-handed positron
\begin{equation}
	|i\rangle_I = |R\rangle_A \otimes |L\rangle_{B},
	\label{Eq.3.1}
\end{equation}
we employ the helicity basis ($|0\rangle \equiv |R\rangle$ and $|1\rangle \equiv |L\rangle$). To characterize the scattering energy conveniently, we define the dimensionless momentum parameter $\mu=|\mathbf{p}|/m_e$, where $|\mathbf{p}|$ is the incoming momentum in the COM reference frame and ${m_e}$ is the electron mass. While helicity coincides with chirality in the ultra-relativistic limit ($\mu \to \infty$), the finite fermion mass in the non-relativistic regime ($\mu\sim1$) explicitly induces non-trivial helicity-flip transitions. To avoid the kinematic divergences associated with the $t$-channel forward-scattering pole ($\theta \to 0, 2\pi$), evaluating the tree-level QED $S$-matrix for non-forward scattering angles yields the normalized final pure state
\begin{equation}
	|f\rangle_I = \frac{1}{\mathcal{N}} \sum_{r,s \in {R,L}} \mathcal{M}(RL; rs)|r\rangle_A |s\rangle_B,\label{Eq.3.2}
\end{equation}
where $\mathcal{M}(RL; rs)$ denotes the helicity-dependent scattering amplitudes (the explicit form of which is provided in Appendix \ref{A}), and $\mathcal{N} = \sqrt{\sum_{r,s} |\mathcal{M}(RL; rs)|^2}$ is the normalization factor ensuring proper probability conservation. To evaluate the entropic uncertainty, we select the Pauli operators $\sigma_z$ and $\sigma_x$ as incompatible observables for subsystem $A$. Physically, measuring $\sigma_z$ extracts the longitudinal helicity population, while $\sigma_x$ probes the transverse quantum superposition. 
The unconditioned post-measurement states are formulated via local projection superoperators
	\begin{equation}
		\begin{split} 
			\rho_{\hat{X}}^f &= \sum_{i} (\Pi_{x_i} \otimes \mathbb{I}_B) \rho^f (\Pi{x_i} \otimes \mathbb{I}_B), \\
			\rho_{\hat{Z}}^f &= \sum_{i} (\Pi_{z_i} \otimes \mathbb{I}_B) \rho^f (\Pi{z_i} \otimes \mathbb{I}_B), 
		\end{split}
		\label{Eq.3.3} 
	\end{equation}
	where $\Pi_{k_i} = |k_i\rangle\langle k_i|$ represent the local projection operators onto the respective eigenstates of $\sigma_x$ and $\sigma_z$, and $\mathbb{I}_B$ is the identity operator acting on the spectator positron. 
	
\begin{figure}[t]
    	\centering
    	\includegraphics[width=8.7cm]{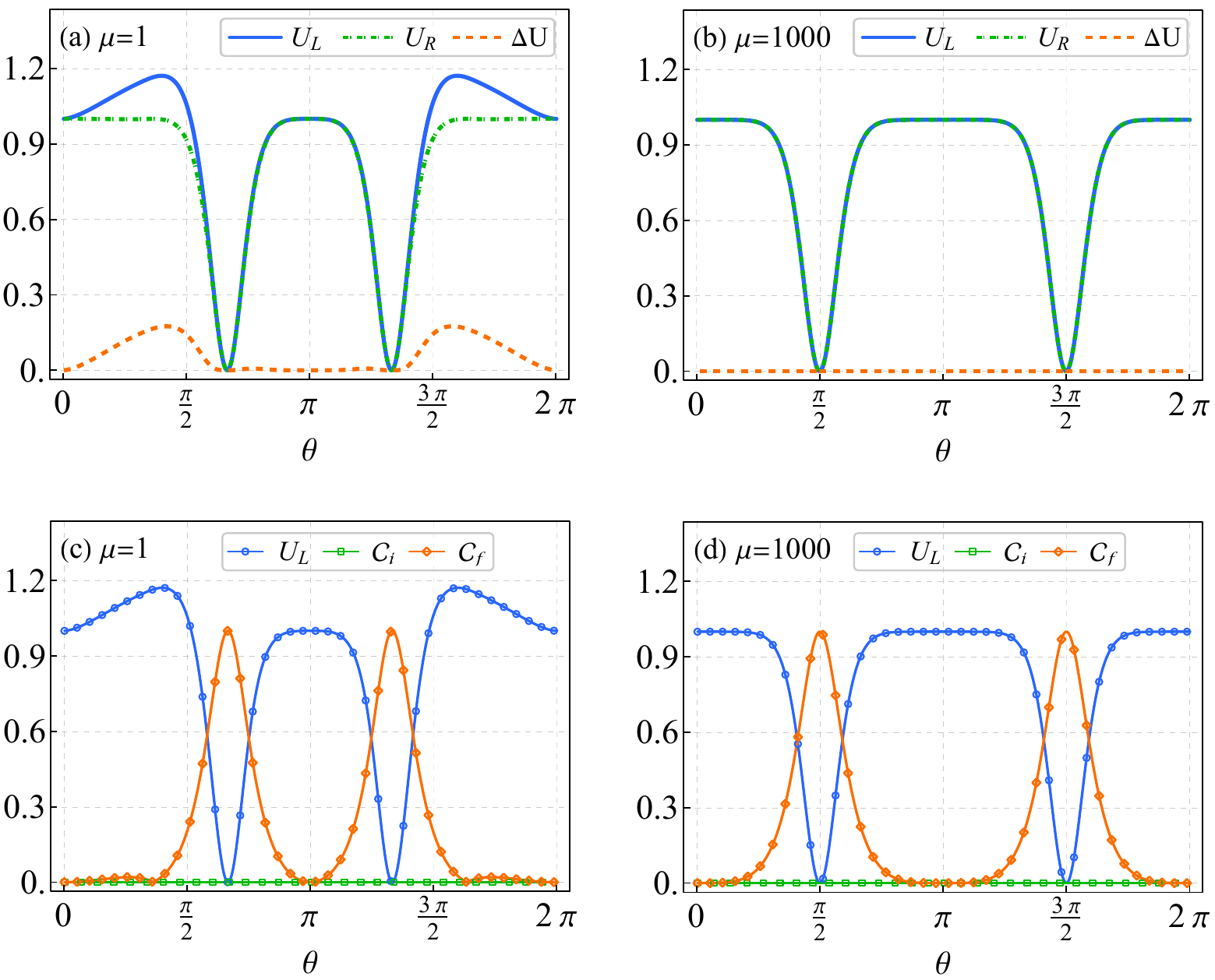}
    	\caption{
    		Dynamical evolution of the entropic uncertainty and quantum entanglement as a function of the scattering angle $\theta$ for the initial factorized state $|i\rangle_I = |R\rangle_A \otimes |L\rangle_B$. Panels (a) and (b) display the entropic uncertainty components—the measured uncertainty $U_L$ (blue solid curves), the theoretical lower bound $U_R$ (green dash-dotted curves), and their tightness difference $\Delta U$ (orange dashed curves)—in the non-relativistic regime ($\mu=1$) and the ultra-relativistic limit ($\mu=1000$), respectively. Panels (c) and (d) depict the strict anti-correlation between the actual entropic uncertainty $U_L$ (blue curves with open circles) and the dynamically generated entanglement $C_f$ (orange curves with open diamonds) across the corresponding momentum regimes. The initial entanglement $C_i$ (green lines with open squares) strictly remains zero.}
    	\label{Fig1}
    \end{figure}

	Combining Eqs. (\ref{Eq.9}), (\ref{Eq.11}) and (\ref{Eq.3.3}), 
	the dynamical evolution of entropic uncertainty and systemic entanglement for the initial separable state is presented in Fig. \ref{Fig1}. Given that the tree-level S-matrix preserves the purity of the bipartite state, the entropic uncertainty lower bound simplifies exactly to $U_R = 1 - S(\rho_A)$. 
    In the non-relativistic regime ($\mu=1$) where  the electron mass is prominent, as depicted in Fig. \ref{Fig1}\textcolor{blue}{(a)}, the measured uncertainty $U_L$ generally exceeds the theoretical lower bound $U_R$, resulting in a non-zero tightness gap ($\Delta U > 0$) across most scattering angles.
    From a quantum field theory perspective, this deviation stems from mass-induced chirality mixing. The non-negligible mass term dynamically activates helicity-flip transition amplitudes [e.g., $\mathcal{M}(RL;RR)$], which inherently possess an asymmetric angular dependence via the $\cot(\theta/2)$ factor. The coherent superposition of these non-vanishing flip amplitudes with the helicity-conserving channels prevents the post-scattering state from magnifying its quantum memory efficiency, leaving the uncertainty bound unsaturated except at specific symmetry points.  Notably, since the system is prepared in a basic separable initial state, the absence of initial coherent superposition prevents the generation of asymmetric interference cross-terms, ensuring that the overall kinematic profile remains strictly symmetric with respect to the central scattering angle $\theta = \pi$. 
    
Conversely, in the ultra-relativistic limit ($\mu=1000$) shown in  Figs. \ref{Fig1}\textcolor{blue}{(b)} and \ref{Fig1}\textcolor{blue}{(d)}, chiral symmetry is effectively restored ($m_{e} \to 0$), forcefully suppressing all single-helicity-flip transitions. Because the final state is strictly confined to a coherent superposition of only the helicity-conserving channels ($|RL\rangle$ and $|LR\rangle$), the system remains a perfect bipartite pure state without any informational leakage. Consequently, the entropic uncertainty bound becomes perfectly saturated ($U_L = U_R$, $\Delta U \equiv 0$) across all scattering angles. 
Strikingly, the entanglement profile in Fig. \ref{Fig1}\textcolor{blue}{(d)} exhibits a needle-like kinematic: the generated entanglement $C_f$ spikes to its theoretical maximum strictly at the scattering angles ($\theta = \pi/2$ and $3\pi/2$), while remaining strongly suppressed elsewhere.  This extreme angular localization is driven by the propagator pole within the $t$-channel Feynman diagram. At almost all scattering angles away from these exact  directions, the $t$-channel amplitude—governed by the photon propagator $1/(1-\cos\theta)$—singularly dominates the relatively constant $s$-channel annihilation process. This overwhelming dominance restricts the system to a highly asymmetric superposition, driving the scattered fermions into a nearly separable configuration heavily dominated by the $|RL\rangle$ component. Only at exactly $\theta = \pi/2$ ($\cos\theta = 0$) is the singular behavior neutralized, allowing the $s$-channel and $t$-channel amplitudes to achieve a perfect 1:1 dynamical equalization. This exact equivalence drives the fermions into a maximally entangled Bell state $\frac{1}{\sqrt{2}}(|RL\rangle + |LR\rangle)$.
Furthermore, across both momentum regimes, Figs. \ref{Fig1}\textcolor{blue}{(c)} and \ref{Fig1}\textcolor{blue}{(d)} reveal a rigorous, perfect anti-correlation between the entropic uncertainty $U_L$ and the quantum entanglement $C_f$. This perfect anti-correlation physically confirms the validity of the quantum-memory paradigm: as the scattering dynamics generate entanglement, the spectator positron perfectly extracts the 'which-helicity' information of the electron, thereby dynamically lowering the uncertainty bound. 
This mechanism is most starkly evident at the exact backward scattering angle ($\theta = \pi$). At this limit, the kinematic angular factor embedded in the $t$-channel numerator inherently vanishes, i.e., $\lim_{\theta \to \pi} \mathcal{M}_t \propto (1+\cos\theta) \to 0$. With the $t$-channel contribution completely suppressed by this kinematic cancellation, the system collapses deterministically into the $s$-channel dominated, fully separable product state $|f\rangle_{\theta=\pi}=|LR\rangle$. Consequently, as the entanglement is completely depleted ($C_f=0$), the quantum memory capacity is nullified, driving the entropic uncertainty to its absolute maximum ($U_L=1.0$)—a kinematic behavior that remains analytically invariant across both $\mu=1$ and $\mu=1000$ scales.
\subsection{Separable states with local coherence}\label{IIIB}
	To further uncover the role of initial quantum coherence in relativistic scattering, we extend our analysis  to a separable states with local coherence 
	\begin{equation}\label{Eq.3.4}
		\begin{split}  	  
			|i\rangle_\text{II}= & \left(\cos \alpha|R\rangle_{A}+e^{i \xi} \sin \alpha|L\rangle_{A}\right) \otimes \\
			& \left(\cos \beta|R\rangle_{B}\right.\left.+e^{i \eta} \sin \beta|L\rangle_{B}\right),
		\end{split}
	\end{equation}
	the resultant final state can be expressed as
	\begin{equation}\label{Eq.3.5}
		\begin{split}  	  
			|f\rangle_\text{II}
			& =\sum_{r, s}\left[\cos \alpha \cos \beta \mathcal{M}(R R ; r s)|r\rangle_{A}|s\rangle_{B}\right. \\
			& +e^{i \eta} \cos \alpha \sin \beta \mathcal{M}(R L ; r s)|r\rangle_{A}|s\rangle_{B} \\
			& +e^{i \xi} \sin \alpha \cos \beta \mathcal{M}(L R ; r s)|r\rangle_{A}|s\rangle_{B} \\
			& \left. +e^{i(\xi+\eta)} \sin \alpha \sin \beta \mathcal{M}(L L ; r s)|r\rangle_{A}|s\rangle_{B}\right] .
		\end{split}
	\end{equation}
	
\begin{figure}[t]
	\centering
	\includegraphics[width=8.7cm]{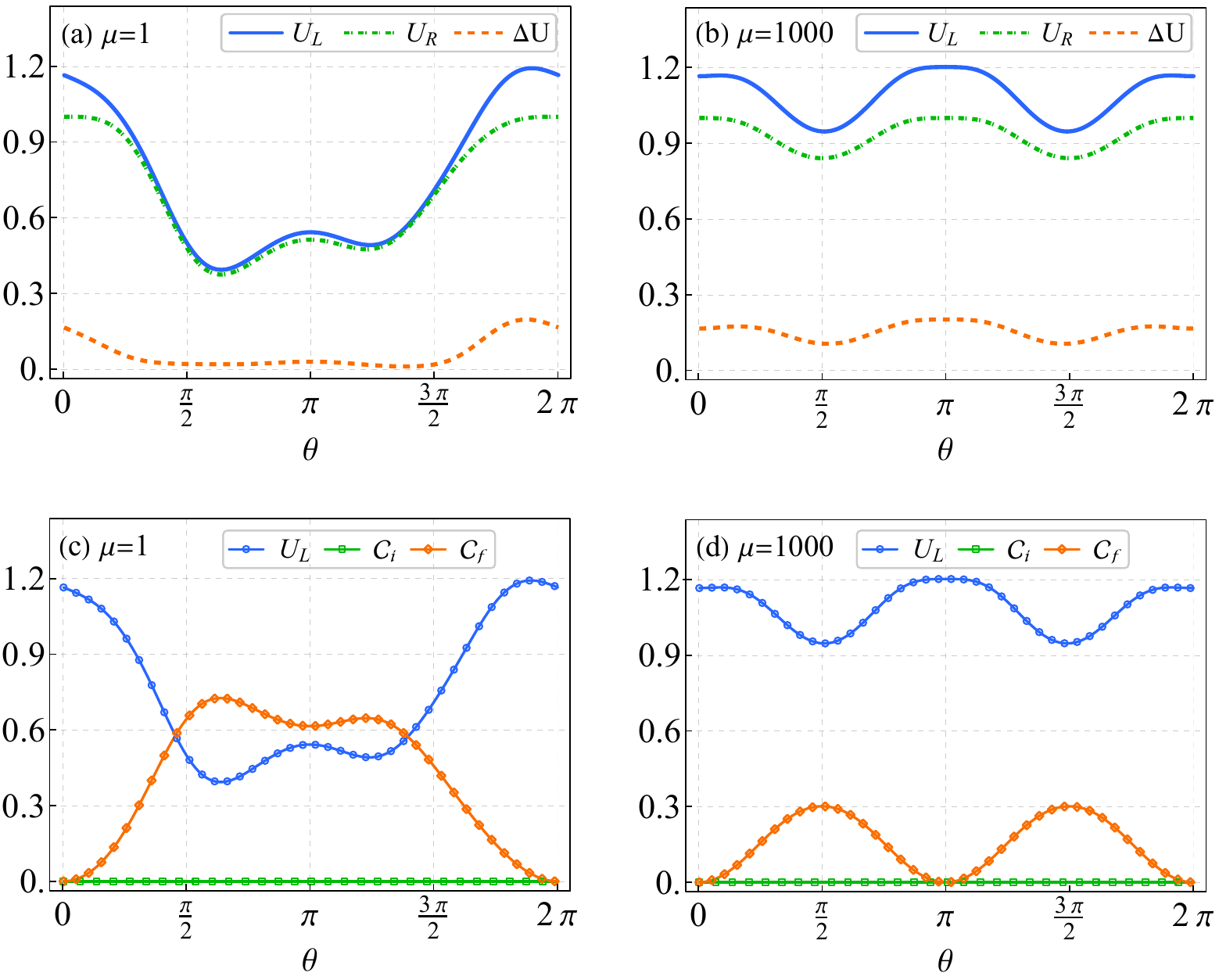}
	\caption{Dynamical evolution of the entropic uncertainty and quantum entanglement as a function of the scattering angle $\theta$ for the separable states with local coherence, parameterized by $\alpha=\pi/3, \beta=3\pi/8$ and vanishing relative phases ($\eta = \xi = 0$). Note the kinematic asymmetry with respect to the scattering angle $\theta=\pi$ in the non-relativistic regime [panels (a) and (c)], which is distinctly restored to a symmetric profile in the ultra-relativistic limit [panels (b) and (d)] due to the rigorous recovery of QED chiral symmetry. The curve and marker styles are identical to those defined in Fig.\ref{Fig1}.}
	\label{Fig2}
\end{figure}
For convenience, the normalization factor has been omitted here temporarily. The evolution of  entropic uncertainty and entanglement is displayed in Fig. \ref{Fig2}. 
{Unlike the basic factorized incoming state, this configuration inherently possesses local coherent superpositions of helicity states prior to the collision, which profoundly reshapes the post-scattering quantum resources. While the fundamental scattering mechanisms—specifically, the mass-induced helicity flips at $\mu=1$ and the chiral symmetry recovery at $\mu=1000$—parallel those detailed in Fig. \ref{Fig1}, the pre-existing initial coherence introduces distinct structural changes:
First, in the non-relativistic  regime, the initial coherent superposition enables a structural cross-interference between the helicity-flipping and helicity-conserving channels. As a result, the geometric symmetry about $\theta=\pi$ is severely broken in Figs. \ref{Fig2}\textcolor{blue}{(a)} and \ref{Fig2}\textcolor{blue}{(c)}. This interference also distributes non-trivial quantum resources across a broader range of scattering angles, driving a remarkable kinematic broadening of the entanglement profile. 
Second, the initial coherence decisively constrains the absolute capacity of entanglement generation. In contrast to Fig.~\ref{Fig1}, where the perfect amplitude equalization at $\theta = \pi/2$ yields a maximally entangled Bell state ($C_f = 1.0$), the non-uniform initial probability weights inherently disrupt this optimal balance, severely capping the entanglement peak well below unity ($C_f \approx 0.3$) even in the ultra-relativistic limit [Fig.~\ref{Fig2}\textcolor{blue}{(d)}].
Due to this dynamical detuning, the total conditional uncertainty $U_L$ fails to fully saturate its lower bound, leaving a persistent tightness gap ($\Delta U > 0$) across all scattering angles even when chiral symmetry is restored [Fig. \ref{Fig2}\textcolor{blue}{(b)}].
Nevertheless, the strict geometric anti-correlation between $C_f$ and $U_L$ remains universally invariant across all conditions, further verifying the structural robustness of the quantum-memory framework.}
\begin{figure}[t]
	\centering
	\includegraphics[width=8.7cm]{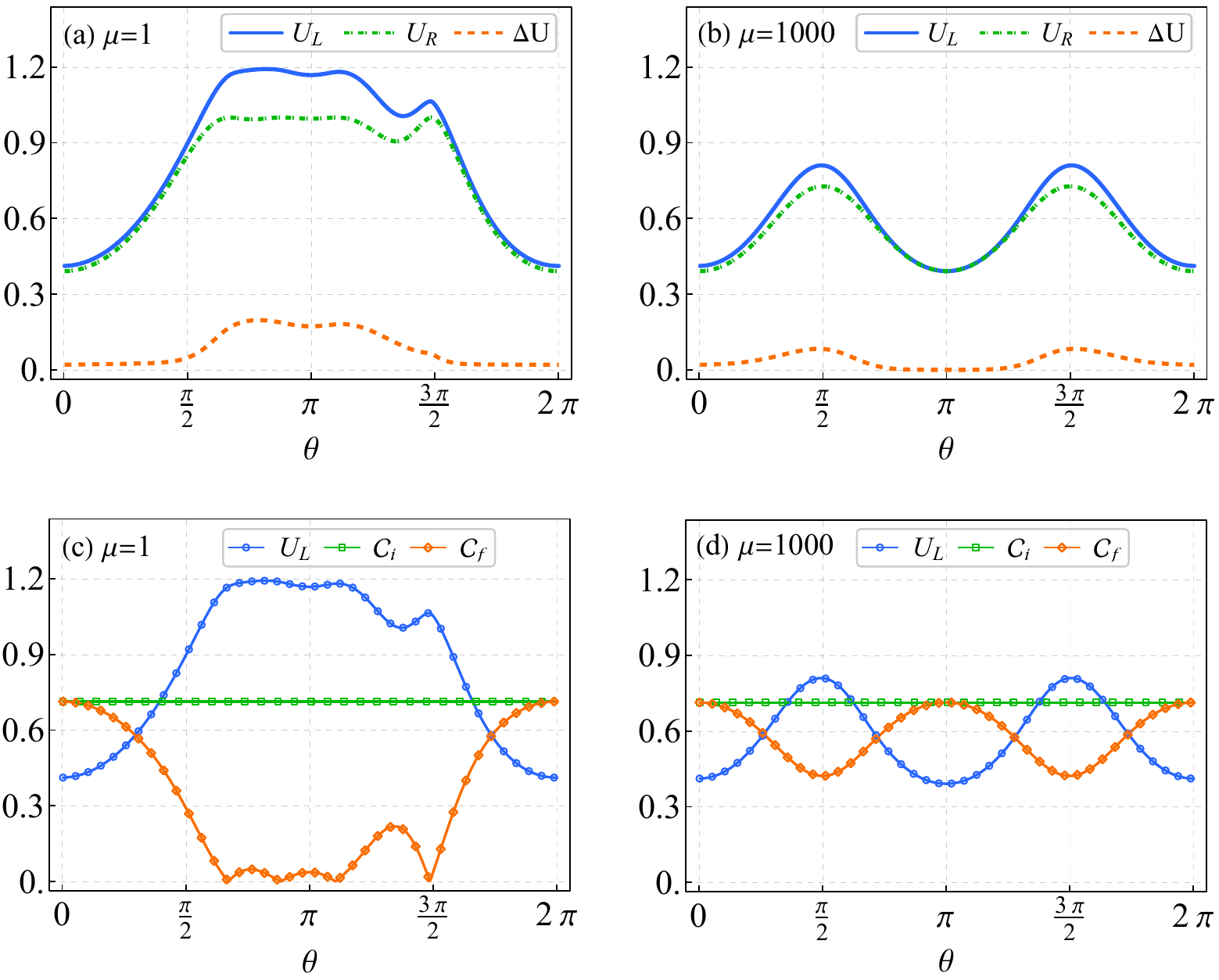}
	\caption{Dynamical evolution of the entropic uncertainty and quantum entanglement as a function of the scattering angle $\theta$ for a generic entanglement-depleting initial state. The initial pure state is parameterized by $\{\alpha=\pi/3, \beta=\pi/5, \chi=0\}$ and vanishing relative phases ($\xi=\eta=\tau=0$). The initial entanglement $C_i \approx 0.72$ is denoted by the green lines with open squares. The universal suppression of the post-scattering entanglement $C_f$ (red curves with open diamonds) below $C_i$ highlights the scattering-induced disentanglement driven by destructive coherent interference. The curve styles for the uncertainty components match those detailed in Fig. 1.}
	\label{Fig3}
\end{figure}
\subsection{General entangled states and scattering-induced disentanglement}\label{IIIC}
	Having established the constructive generation of quantum correlations from initial separable states in the preceding sections, we now proceed to the most general scattering scenario. A fundamental question arises: does the tree-level QED interaction universally act as an entanglement amplifier, or can scattering kinematics actively deplete pre-existing non-local resources ? To address this, we consider an initial electron-positron pair that already shares an arbitrary degree of bipartite entanglement. Such a generic two-qubit pure incoming state is parameterized by three probability amplitude angles ($\alpha, \beta, \chi$) and three relative phases ($\xi, \eta, \tau$), expressed as
	\begin{equation}\label{Eq.3.6}
		\begin{split} 
			|i\rangle_{\text{III}}&=\cos\alpha|R\rangle_A|R\rangle_B+e^{i\xi}\sin\alpha\cos\beta|R\rangle_A|L\rangle_B \\
			&+e^{i\eta}\sin\alpha\sin\beta\cos\chi|L\rangle_A|R\rangle_B\\
			&+e^{i\tau}\sin\alpha\sin\beta\sin\chi|L\rangle_A|L\rangle_B.
		\end{split}
	\end{equation}
	Through unitary evolution governed by the QED $S$-matrix, the transition amplitudes of all helicity channels coherently superimpose. The corresponding post-scattering final state $|f\rangle_{\text{III}}$ is rigorously formulated as
	\begin{equation}
		\begin{split}  	  
			|f\rangle_{\text{III}}&=\sum_{r,s} \Big[ \cos\alpha\mathcal{M}(RR;rs) \\
			&+e^{i\xi}\sin\alpha\cos\beta\mathcal{M}(RL;rs) \\
			&+e^{i\eta}\sin\alpha\sin\beta\cos\chi\mathcal{M}(LR;rs) \\
			&+e^{i\tau}\sin\alpha\sin\beta\sin\chi\mathcal{M}(LL;rs) \Big] |r\rangle_{A}|s\rangle_{B}.
		\end{split}\label{Eq.3.7}
	\end{equation}

	 Analytically, the post-scattering entanglement landscape is highly sensitive to the initial quantum superposition. Depending on the coherent interference between the $s$- and $t$-channel amplitudes, the generic initial states can phenomenologically manifest as entanglement-enhancing (where correlations are further amplified) or entanglement-depleting (where pre-existing entanglement monotonically degrades). Since the constructive generation of entanglement has been comprehensively demonstrated in Sections \ref{IIIA} and \ref{IIIB}, we strategically restrict our focus in this section to a representative entanglement-depleting configuration. By selecting the specific parameter set $\{\alpha =\pi/3, \beta =\pi/5, \chi=0\}$ alongside vanishing relative phases ($\xi=\eta=\tau=0$), we isolate a kinematic scenario that severely suppresses non-local correlations. This specific selection allows us to critically investigate the destructive interference mechanisms of relativistic scattering, demonstrating how the QED $S$-matrix can dynamically deplete pre-existing entanglement and drive a concomitant surge in the entropic uncertainty bound.

     The dynamical evolution of quantum resources for a general entangled incoming state ($\alpha =\pi/3, \beta =\pi/5$) is comprehensively visualized in Fig. \ref{Fig3}. The most distinctive feature, evident in both Figs. \ref{Fig3}\textcolor{blue}{(c)} and \ref{Fig3}\textcolor{blue}{(d)}, is that the post-scattering entanglement $C_f$ (red curves) is stringently suppressed below its initial value $C_i \approx 0.72$ (green lines) across nearly the entire angular range. This widespread depletion demonstrates the kinematic fragility of pre-existing quantum correlations under QED interactions, where the specific coherent superposition of the initial state drives intense destructive interference that actively scrambles non-local resources into localized single-particle attributes.
     While the underlying kinematic symmetries—specifically,  
     the mass-induced geometric asymmetry at $\mu=1$ and its restoration at $\mu=1000$—follow the same physics detailed in Figs. \ref{Fig1} and \ref{Fig2}, this general initial state introduces novel dynamical phenomena:
     In the low-energy regime [Figs. \ref{Fig3}\textcolor{blue}{(a)}  and \ref{Fig3}\textcolor{blue}{(c)} ], the mass-induced cross-interference between helicity-flip and helicity-conserving channels severely breaks the geometric symmetry about $\theta = \pi$. Remarkably, as shown in Fig. \ref{Fig3}\textcolor{blue}{(c)} , this destructive interference is so profound that the emergent entanglement $C_f$ is driven to absolute zero at two distinct kinematic configurations: a broad suppression plateau near $\theta \approx 2\pi/3$ and a sharp dip at $\theta = 3\pi/2$. At these specific nodes, the coherent cancellation completely severs the bipartite correlation, collapsing the system into fully separable product states and consequently driving the entropic uncertainty $U_L$ (blue curve) to its local maxima.

     In the ultra-relativistic regime, conversely, when chiral symmetry is restored [Figs. \ref{Fig3}\textcolor{blue}{(b)}  and \ref{Fig3}\textcolor{blue}{(d)} ], the profile regains strict symmetry with respect to $\theta=\pi$. Here, the destructive interference reaches its peak at the scattering angle ($\theta=\pi/2, 3\pi/2$), where the equalized $s$- and $t$-channel amplitudes severely degrade $C_f$ to its absolute minima. Most notably, a lossless entanglement preservation anomaly emerges strictly at the exact backward scattering angle ($\theta=\pi$), as well as at the extreme forward limits ($\theta \to 0, 2\pi$). At $\theta=\pi$, the helicity selection rules and kinematic constraints allow the scattering matrix to act as a non-depleting rotation that perfectly maps the initial Schmidt rank onto the final state. Consequently, $C_f$ dynamically rebounds to perfectly touch the initial threshold $C_i$, where the actual uncertainty $U_L$ and the lower bound $U_R$ perfectly osculate ($\Delta U = 0$) in Fig. \ref{Fig3}\textcolor{blue}{(b)}, confirming an optimal quantum memory capacity.
    Throughout all variations, the rigid geometric mirroring between $U_L$ and $C_f$ remains universally locked, further establishing the robustness of the quantum-memory-assisted framework under arbitrary QED scattering configurations.
	%
	\begin{figure*}[htbp]
		\centering
		\includegraphics[width=18cm]{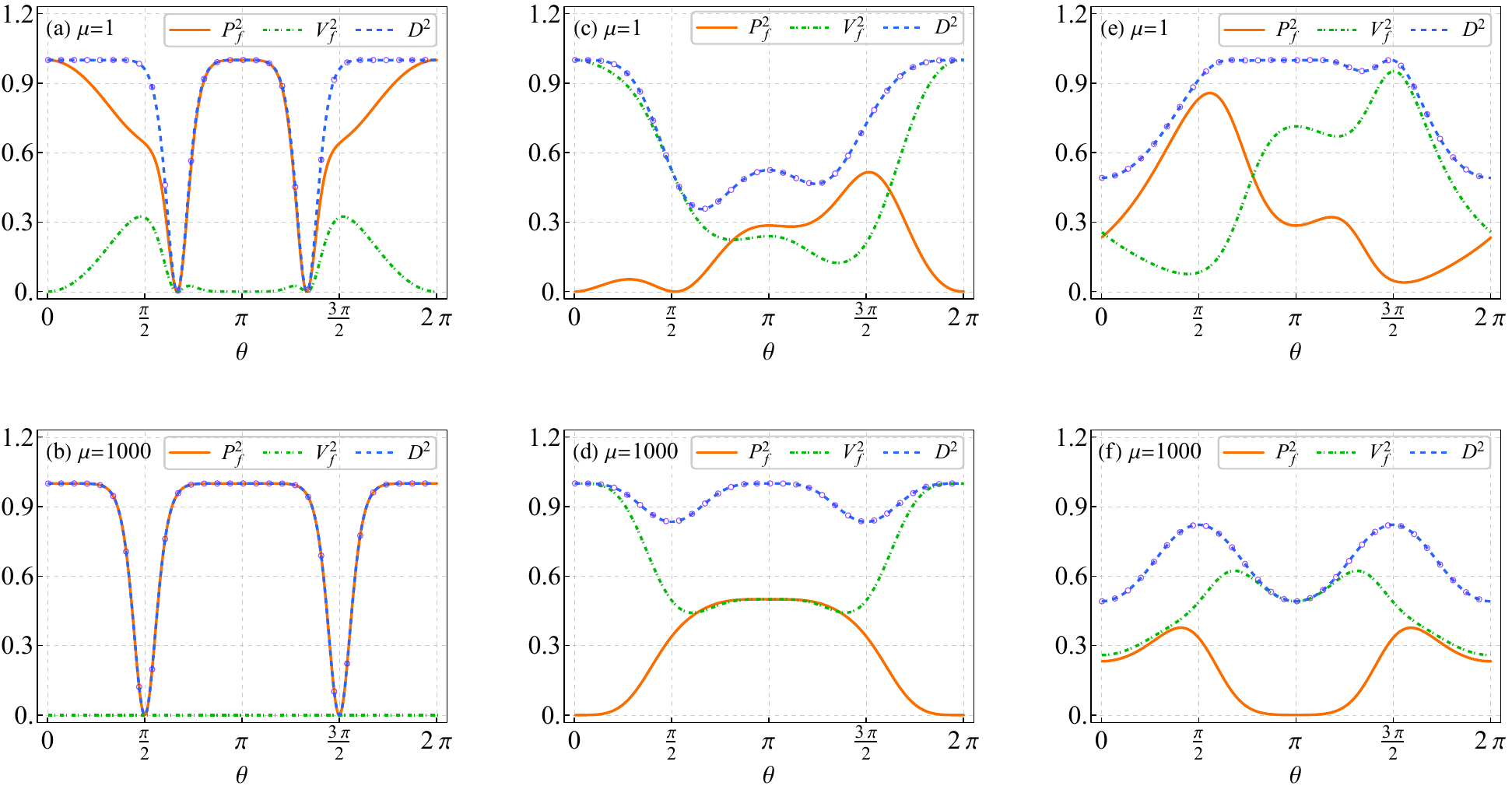}
		\caption{Scattering angle  dependence of the local duality parameters and global quantum coherence. The predictability $P_f^2$ (solid orange curves), wave-like visibility $V_f^2$ (dash-dotted green curves), $P_f^2$+$V_f^2$ (purple circle), and global coherence $D^2$ (dashed blue curves) are plotted for three distinct initial configurations: the basic factorized state (left column, panels (a) and (b)), the separable states with local coherence is parameterized by $\{\alpha=\pi/4, \beta=3\pi/8, \xi=\eta=0\}$ (middle column, panels (c) and (d)),  and the general entangled states is parameterized by $\{\alpha=\pi/3, \beta=\pi/5, \chi=0\, \xi=\eta=\tau=0\}$ (right column, panels (e) and (f)). The top row ($\mu=1$) and bottom row ($\mu=1000$) correspond to the non-relativistic regime and the ultra-relativistic limit, respectively.}
		\label{Fig4}
	\end{figure*}
\section{Equivalence relation between duality and quantum coherence}\label{IV}
{For any two-qubit pure state generated from the scattering, the post-scattering state of the system can be parameterized as
	\begin{equation}\label{Eq.4.1}
		|\varphi\rangle_{f} =\mathcal{A}|RR\rangle+\mathcal{B}|RL\rangle+\mathcal{C}|LR\rangle+\mathcal{D}|LL\rangle,
	\end{equation}
	where the complex probability amplitudes satisfy the normalization condition $|\mathcal{A}|^2 + |\mathcal{B}|^2 + |\mathcal{C}|^2 + |\mathcal{D}|^2 = 1$. To evaluate the local wave-particle properties of electron $A$, one must perform a partial trace over the kinematic degrees of freedom of the outgoing positron $B$. This operation yields the reduced density matrix $\rho_A$, wherein the diagonal elements govern the dynamical which-helicity predictability $P_A$, and the off-diagonal elements dictate the local interference visibility $V_A$. Using Eq.~(\ref{Eq.12}), we obtain:
	\begin{equation}\label{Eq.4.2}
		\begin{split}
			P_{A}^2 &= (1 - 2(|\mathcal{C}|^2 + |\mathcal{D}|^2))^2,\\
			V_{A}^2 &= 4|\mathcal{A}\mathcal{C}^* + \mathcal{B}\mathcal{D}^*|^2.
		\end{split}
	\end{equation}
		Concurrently, according to Eq. (\ref{Eq.13}) evaluating the global bipartite coherence $D_{AB}^2$, which inherently quantifies the purity and the total non-classical correlation of the superposed scattering amplitudes
\begin{equation}\label{Eq.3.4}
	D_{AB}^2 = 1 - 4|\mathcal{A}\mathcal{D} - \mathcal{B}\mathcal{C}|^2.
\end{equation}
By factoring in the normalization constraint of the scattering matrix, the algebraic structures of these quantities perfectly coincide, leading directly to the following fundamental equivalence
\begin{equation}\label{Eq.4.4}
P_{A}^2+V_{A}^2=D_{AB}^2.
\end{equation}
By virtue of the Schmidt decomposition for bipartite pure states, the reduced density matrices $\rho_A$ and $\rho_B$ share identical non-zero eigenvalues [i.e., $\text{Tr}(\rho_A^2) = \text{Tr}(\rho_B^2)$]. Consequently, this equivalence relation is universally symmetric with respect to either subsystem, strictly guaranteeing that $P_{B}^2+V_{B}^2=D_{AB}^2$.

Fig. (\ref{Fig4}) illustrates the dynamical evolution of the local duality parameters and global coherence across different initial states and energy scales. The most fundamental observation is the exact coincidence of $D^2$ (dashed blue curves) with $P_f^2 + V_f^2$ (purple circle) under all kinematic conditions. This result rigorously validates the equivalence relation $P_A^2 + V_A^2 = D_{AB}^2$ derived in Eq. (\ref{Eq.4.4}), establishing that the quantum resource conservation law remains inviolable throughout the unitary QED scattering process. The specific dynamical features across the subplots are governed by the interplay between initial local coherence and chiral symmetry. 
 For the non-relativistic regime [Figs. \ref{Fig4}\textcolor{blue}{(a)}, \ref{Fig4}\textcolor{blue}{(c)}, and \ref{Fig4}\textcolor{blue}{(e)}], the finite fermion mass dynamically activates single-helicity-flip transitions. For the basic factorized state [Fig. \ref{Fig4}\textcolor{blue}{(a)}], the lack of initial coherence precludes cross-interference; thus, the kinematic profile remains strictly symmetric with respect to scattering angle $\theta = \pi$ while $D^2$ is partitioned between $P_f^2$ and $V_f^2$. However, for the generalized superpositions [Figs. \ref{Fig4}\textcolor{blue}{(c)} and \ref{Fig4}\textcolor{blue}{(e)}], 
 the pre-existing initial coherence structurally cross-interferes with these mass-induced flip channels. 
 This interference directly exposes the asymmetric angular dependence of the flip amplitudes. Consequently, the geometric symmetry about $\theta = \pi$ is severely broken, and the local wave-particle attributes—along with the global coherence envelope—are dynamically redistributed and distorted across all scattering angles.
 
Conversely, in the ultrarelativistic regime [Figs.~\ref{Fig4}\textcolor{blue}{(b)}, \ref{Fig4}\textcolor{blue}{(d)}, and \ref{Fig4}\textcolor{blue}{(f)}], the rigorous recovery of chiral symmetry suppresses the single-helicity-flip amplitudes, forcefully ensuring that the kinematic profile remains strictly symmetric with respect to scattering angle $\theta = \pi$.
For the basic factorized state [Fig.~\ref{Fig4}\textcolor{blue}{(b)}], the chiral symmetry strictly confines the system to the helicity-conserving channels. Coupled with the complete absence of initial coherence, this suppresses all off-diagonal elements in the reduced density matrix, forcing the local visibility to vanish globally ($V_f^2 \equiv 0$). 
For states with pre-existing local coherence [Figs.~\ref{Fig4}\textcolor{blue}{(d)} and \ref{Fig4}\textcolor{blue}{(f)}], however, the initial superposition guarantees $V_f^2 > 0$. 
Notably, at the exact backward scattering angle $\theta = \pi$, these coherent initial configurations yield two distinct, mathematically elegant duality limits. 
For the separable state with local superposition [Fig.~\ref{Fig4}\textcolor{blue}{(d)}], the QED interaction acts effectively as a 50:50 quantum beam-splitter, exactly equipartitioning the global coherence into the wave and particle domains ($P_f^2 = V_f^2 = \frac{1}{2}D^2$). 
This precise halfway split indicates a state of perfect duality balance. 
In stark contrast, for the general entangled incoming state [Fig.~\ref{Fig4}\textcolor{blue}{(f)}], the local predictability is strictly eradicated ($P_f^2 \to 0$) at this same kinematic node. Consequently, the available global coherence is entirely channeled into local chiral interference ($V_f^2 = D^2$), definitively stripping the fermion of its classical particle identity. Meanwhile, at the scattering angle $\theta = \pi/2, 3\pi/2$, the global coherence reaches its maxima, which concurrently enhances both the predictability and the wave-like visibility.

	\section{Kinematic Trade-off Between Local Duality and Non-locality }\label{V}
	\begin{figure*}
		\begin{minipage}{1\textwidth}
			\centering
			\subfigure{\includegraphics[width=18cm]{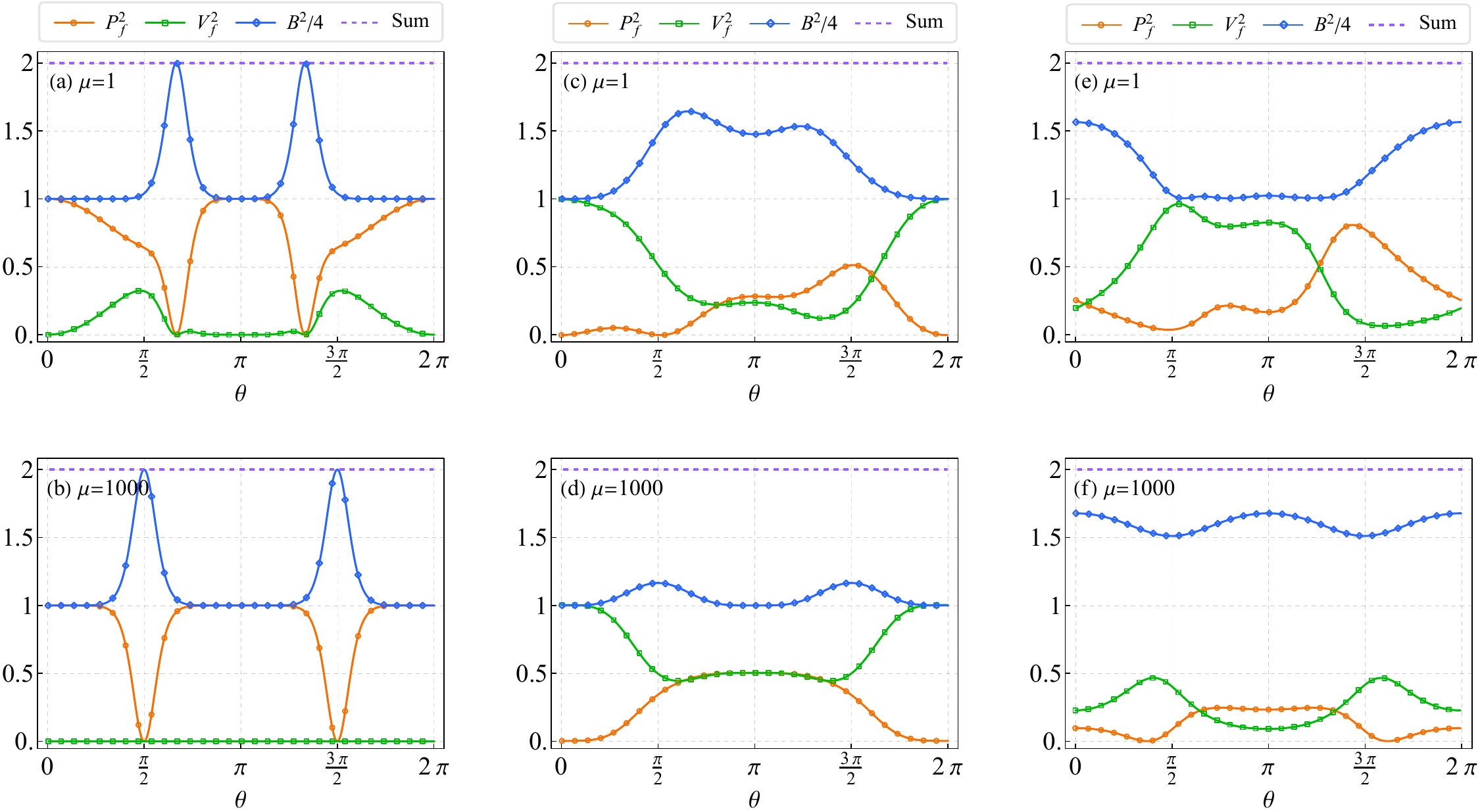}} 
		\end{minipage}\hfill
		\caption{Scattering angle dependence of local wave-particle duality and non-local Bell violation. The local predictability $P_f^2$ (orange curves), the wave-like visibility $V_f^2$ (green curves), the scaled CHSH expectation value $B^2/4$ (blue curves), and their sum (dashed purple curves) are plotted for the basic factorized state  (left column), the separable states with local coherence is parameterized by $\{\alpha=\pi/4, \beta=3\pi/8, \xi=\eta=0\}$ (middle column), and the general entangled states is parameterized by $\{\alpha=\pi/2, \beta=\pi/3, \chi=\pi/6, \xi=\eta=\tau=0\}$ (right column). The top ($\mu=1$) and bottom ($\mu=1000$) rows correspond to the non-relativistic and ultra-relativistic regimes, respectively.}
		\label{Fig5}
	\end{figure*}

	  Building upon the equivalence established in Section \ref{IV}, we now investigate the fundamental trade-off between the local wave-particle duality and the violation of local realism. For a bipartite system, classical local hidden-variable theories are bounded by the Clauser-Horne-Shimony-Holt (CHSH) inequality, $|B_{\text{CHSH}}| \le 2$ \cite{Bell3.Clauser.PhysRevLett.23.880}. However, for the post-scattering pure state $|\varphi\rangle_f$ [Eq. (\ref{Eq.4.1})], the maximum expectation value of the CHSH operator, $B_{\max}$, can exceed this classical limit. According to the Horodecki criterion \cite{Bell4.1PhysRevLett.89.170401}, $B_{\max}$ is a strict monotonic function of the bipartite Concurrence $C_f$
		\begin{equation}\label{Eq.5.1}
			B_{\max} = 2\sqrt{1 + C_f^2}.
		\end{equation}
		To map how this non-local resource restricts the local physical identity of the individual fermions, we substitute $C_f^2 = (B_{\max}^2 / 4) - 1$ into the complete complementarity relation Eq. (\ref{Eq.10}). This rigorously yields the exact analytical relationship between local duality and Bell's inequality
		\begin{equation}\label{Eq.5.2}
			P_A^2 + V_A^2 + \frac{B_{\max}^2}{4} = 2,
		\end{equation}
	  the equation establishes a dynamic trade-off relationship between local kinematic reality and non-local relativistic correlations. The sum $P_A^2 + V_A^2$ encapsulates the total accessible local identity of the electron.

	Fig. \ref{Fig5} comprehensively visualizes the dynamical interplay among the local wave-particle attributes ($P_f^2$, $V_f^2$) and the bipartite non-local correlation ($B^2/4$). The most fundamental observation is the exact coincidence of their total sum at $2$ under all kinematic conditions.
	Since the underlying scattering mechanisms—namely, the mass-induced cross-interference at $\mu=1$ and the chiral symmetry recovery at $\mu=1000$—are identical to those detailed in Fig. \ref{Fig4}, we focus here strictly on their consequences for the Bell violation. 
	For the non-relativistic regime,  [Figs. \ref{Fig5}\textcolor{blue}{(a)}, \ref{Fig5}\textcolor{blue}{(c)}, and \ref{Fig5}\textcolor{blue}{(e)}], the asymmetric redistribution of local wave-particle duality severely restricts the generation of non-locality. Most strikingly, in the general entangled states [Fig. \ref{Fig5}\textcolor{blue}{(e)}], the mass-induced constructive interference triggers a surge in local visibility ($V_f^2 \to 1$). Governed by the strict constraint of Eq.~(\ref{Eq.5.2}), this forced localization of coherence mathematically compels a severe kinematic dampening of the non-local correlations. Consequently, in the scattering angle range $\theta \in [\pi/2, 3\pi/2]$, $B^2/4$ is stringently suppressed to the classical local realism threshold.
	Conversely, in the ultra-relativistic regime  [Figs. \ref{Fig5}\textcolor{blue}{(b)}, \ref{Fig5}\textcolor{blue}{(d)}, and \ref{Fig5}\textcolor{blue}{(f)}], the restoration of strict geometric symmetry about $\theta = \pi$ shapes a more structured resource distribution. For the basic factorized state [Fig. \ref{Fig5}\textcolor{blue}{(b)}], the complete suppression of visibility ($V_f^2 = 0$) allows the predictability to strictly vanish at the transverse scattering angles ($\theta = \pi/2, 3\pi/2$). This total deprivation of local attributes uniquely drives the Bell violation to the absolute quantum limit (the Cirel'son bound, $B^2/4 \to 2$). However, For the states with pre-existing coherence [Figs. \ref{Fig5}\textcolor{blue}{(d)} and \ref{Fig5}\textcolor{blue}{(f)}], the pre-existing local coherence acts as a persistent resource sink ($V_f^2 > 0$). Under the constraint of a constant sum of $2$, this inherent wave-like attribute effectively disrupts the optimal balance of scattering amplitudes, thereby dampening the maximal non-local Bell correlations.

	\section{DISCUSSIONS AND CONCLUSIONS} \label{VI}
	In summary, we have investigated the dynamic generation, conversion, and trade-offs of quantum resources in tree-level Bhabha scattering. Our results illustrate how the QED $S$-matrix redistributes local wave-particle attributes and global non-classical correlations under different kinematic regimes. The main conclusions of this work are threefold: 
	First, we observed a strict anti-correlation between the entropic uncertainty $U_L$ and the post-scattering entanglement $C_f$ across different initial states. In the non-relativistic regime, the mass-induced single-helicity-flip channels break the geometric symmetry about $\theta = \pi$, leading to angle-dependent entanglement depletion. In the ultra-relativistic limit, chiral symmetry is restored, and the scattering profile regains symmetry about $\theta = \pi$. Notably, at this exact backward scattering angle, the process behaves as a non-depleting local unitary operation, which recovers the initial entanglement threshold and strictly saturates the entropic uncertainty lower bound ($\Delta U = 0$). 
	Second, we analytically established the equivalence relation between the local wave-particle duality and the global bipartite coherence, expressed as $P_A^2 + V_A^2 = D_{AB}^2$. This identity holds throughout the unitary scattering process. 
   Under the high-energy limit, the exact backward scattering ($\theta = \pi$) highlights two limiting cases contingent upon specific initial parameter configurations: an equal distribution of wave and particle attributes ($P_f^2 = V_f^2 = D^2/2$) for separable states with local coherence, and a complete conversion of global coherence into local interference ($V_f^2 = D^2, P_f^2 \to 0$) for generally entangled incoming states.
	Third, our analysis reveals a trade-off between local wave-particle duality and Bell nonlocality. Under the constraint of a constant total sum, pre-existing local visibility disrupts the optimal balance of scattering amplitudes. Consequently, the maximal non-local correlation is diminished. 
	
	Operationally, the resource-theoretic frameworks discussed herein indicate that global non-classical correlations and coherence properties can be evaluated via local kinematic observables. Extending this theoretical framework to include higher-order QED radiative corrections or asymmetric initial beam polarizations would be a logical next step. Such studies could further characterize quantum resource dynamics and potential decoherence mechanisms in fundamental high-energy scattering processes. Ultimately, these insights may provide valuable theoretical guidance for the development of QED-based quantum information processing protocols.

	\begin{acknowledgements}
	This work was supported by the National Natural Science Foundation of China (Grants No. 12475009 and No. 12075001), Anhui Province Science and Technology Innovation Project (Grant No. 202423r06050004), Anhui Provincial Natural Science Foundation (Grant No. 2508085ZD001), Anhui Provincial University Scientific Research Major Project (Grant No. 2024AH040008), and Anhui Provincial Department of Industry and Information Technology (Grant No. JB24044).
	\end{acknowledgements}
	
	\appendix
	\begin{widetext}
		\section{Bhabha scattering amplitudes}\label{A}
		{The scattering amplitude is calculated in the COM reference frame of particles $A$ and $B$. In the following,  $p_1 = (\omega, 0, 0, \left|\vec{p}\right |)$ and $p_2 = (\omega, 0, 0, - \left |\vec{p}\right |)$ are the incoming 4-momenta that lie along the $z$-axis, while $p_3 = (\omega, \left|\vec{p}\right| \text{sin}\theta, 0, \left|\vec{p}\right| \text{cos}\theta)$ and $p_4 = (\omega, -\left|\vec{p}\right| \text{sin}\theta, 0, -\left |\vec{p}\right | \text{cos}\theta)$ are the outgoing 4-momenta lying along a direction that form an angle $\theta$ with respect to $z$-axis $a, b, r, s$ are the spin indices,
			\begin{equation}
				\mathcal{M}_{\mathrm{Bhabha}}=e^2\left(\bar{v}(b,p_2)\gamma^\mu u(a,p_1)\frac{1}{(p_1+p_2)^2}\bar{u}(r,p_3)\gamma_\mu v(s,p_4)-\bar{v}(b,p_2)\gamma^\mu v(s,p_4)\frac{1}{(p_3-p_1)^2}\bar{u}(r,p_3)\gamma_\mu u(a,p_1)\right).
			\end{equation}	
			Defining $\mu\equiv\frac{|\mathbf{p}|}{m_e}$, where $|\mathbf{p}|$ is the incoming momentum in the COM reference frame and ${m_e}$ is the electron mass, the explicit expressions for the polarized amplitudes are given by 		\begin{equation}\mathcal{M}(RR;RR)=\mathcal{M}(LL;LL)=\frac{(2+11\mu^2+8\mu^4+2\cos\theta+\mu^2\cos2\theta)\csc^2(\frac{\theta}{2})}{4\mu^2(1+\mu^2)},\end{equation}
			\begin{equation}\mathcal{M}(RR;_{LR}^{RL})=-\mathcal{M}(LL;_{LR}^{RL})=-\frac{(1+\mu^2\cos\theta)\cot(\frac{\theta}{2})}{\mu^2\sqrt{1+\mu^2}},\end{equation}
			\begin{equation}\mathcal{M}(RR;LL)=\mathcal{M}(LL;RR)=\frac{1+\mu^2(1+\cos\theta)}{\mu^2(1+\mu^2)},\end{equation}
			\begin{equation}\mathcal{M}(_{LR}^{RL};RR)=-\mathcal{M}(_{LR}^{RL};LL)=\frac{(1+\mu^2\cos\theta)\cot(\frac{\theta}{2})}{\mu^2\sqrt{1+\mu^2}},\end{equation}
			\begin{equation}\mathcal{M}(RL;RL)=\mathcal{M}(LR;LR)=\frac{(1+\mu^2(1+\cos\theta))\cot^2(\frac{\theta}{2})}{\mu^2},\end{equation}
			\begin{equation}\mathcal{M}(RL;LR)=\mathcal{M}(LR;RL)=1-\cos\theta-\frac{1}{\mu^2}.\end{equation} 
		}
	\end{widetext}

	\newpage
	\appendix 
\end{document}